\documentclass[11pt,a4paper]{article}
\usepackage{amsmath}
\usepackage[mathcal]{eucal}
\usepackage{latexsym}
\usepackage{amssymb}
\begin{titlepage}
\title{Instanton representation of Plebanski gravity. Consistency of the initital value constraints under time evolution}
\author{Eyo Eyo Ita III}
\medskip
\input amssym.def
\input amssym.tex

\begin{document}
\maketitle
\bigskip
\centerline{Department of Applied Mathematics and Theoretical Physics} 
\smallskip
\centerline{Centre for Mathematical Sciences, University of Cambridge, Wilberforce Road}
\smallskip
\centerline{Cambridge CB3 0WA, United Kingdom}
\smallskip
\centerline{eei20@cam.ac.uk} 
\bigskip  
                 
\begin{abstract}  
The instanton representation of Plebanski gravity provides as equations of motion a Hodge self-duality condition and a set of `generalized' Maxwell's equations, subject to gravitational degrees of freedom encoded in the initial value constraints of general relativity.  The main result of the present paper will be to prove that this constraint surface is preserved under time evolution.  We carry this out not using the usual Dirac procedure, but rather the Lagrangian equations of motion themsleves.  Finally, we provide a comparison with the Ashtekar formulation to place these results into overall context.
\end{abstract}
\end{titlepage}
 
\section{Introduction}
  
In \cite{EYOITA} a new formulation of general relativity was presented, named the instanton representation of Plebanski gravity.  The basic dynamical variables 
are an $SO(3,C)$ gauge connection $A^a_{\mu}$ and a matrix $\Psi_{ae}$ taking its values in two copies of $SO(3,C)$.\footnote{Index labelling conventions for this paper are that symbols $a,b,\dots$ from the beginiing of the Latin alphabet denote internal $SO(3,C)$ indices while those from the 
middle $i,j,k,\dots$ denote spatial indices.  Both of these sets of indices take takes $1$, $2$ and $3$.  The Greek symbols $\mu,\nu,\dots$ refer to spacetime indices which take values $0,1,2,3$.}  The consequences of the associated 
action $I_{Inst}$ were determined via its equations of motion, which hinge crucially on weak equalities implied by the the initial value constraints.  For these consequences to be self-consistent, the constraint surface must be preserved for all time by the evolution equations.  The present paper will demonstrate that this is indeed the case.  We will not use the usual Hamiltonian formulation for totally constrained systems \cite{DIR}, since we will not make use of any canonical structure implied by $I_{Inst}$.  Rather, we will deduce the time evolution of the dynamical variables directly from the equations of motion of $I_{Inst}$. \par
\indent
Sections 2 and 3 of this paper present the instanton representation action and derive the time evolution of the basic variables.  Sections 4, 5 and 6 demonstrate that the nondynamical equations, referred to as the diffeomorphism, Gauss' law and Hamiltonian constraints, evolve into combinations of the same constraint set.  The result is that the time derivatives of these constraints are weakly equal to zero with no additional constraints generated on the system.  While we do not use the usual Dirac method in this paper, the result is still that the instanton representation is in a sense Dirac consistent.  We will make this inference clearer by comparison with the Ashtekar variables in the discussion section.  On a final note, the terms `diffeomorphism' and `Gauss' law' constraints are used loosely in this paper, in that we have not specified what transformations of the basic variables these constraints generate.  The use of these terms is mainly for notational purposes, due to their counterparts which appear in the Ashtekar variables.

\section{Instanton representation of Plebanski gravity}

The starting action for the instanton representation of Plebanski gravity is given by \cite{EYOITA}
\begin{eqnarray}
\label{OPTION78}
I_{Inst}=\int{dt}\int_{\Sigma}d^3x\Psi_{ae}B^k_e\bigl(F^a_{0i}+\epsilon_{kjm}B^j_aN^m\bigr)\nonumber\\
-iN(\hbox{det}B)^{1/2}\sqrt{\hbox{det}\Psi}\bigl(\Lambda+\hbox{tr}\Psi^{-1}\bigr),
\end{eqnarray}
\noindent
where $N^{\mu}=(N,N^i)$ are the lapse function and shift vector from metric general relativity, and $\Lambda$ is the cosmological constant.  The basic fields are $\Psi_{ae}$ and $A^a_i$, and we action (\ref{OPTION78}) is defined only on configurations restricted to $(\hbox{det}B)\neq{0}$ and $(\hbox{det}\Psi)\neq{0}$.\footnote{The latter case limits the application of our results to spacetimes of Petroc Types I, D and O (See e.g. 
\cite{MACCALLUM} and \cite{PENROSERIND}.}  In the Dirac procedure one refers to $N^{\mu}$ as nondynamical fields, since their velocities do not appear in the action.  While the velocity $\dot{\Psi}_{ae}$ also does not appear, we will distinguish this field from $N^{\mu}$ since the action (\ref{OPTION78}), unlike for the latter, is nonlinear in $\Psi_{ae}$.\par
\indent
The equation of motion for the shift vector $N^i$, the analogue of the Hamilton equation for its conjugate momentum $\Pi_{\vec{N}}$, is given by
\begin{eqnarray}
\label{SHIFT}
{{\delta{I}_{Inst}} \over {\delta{N}^i}}=\epsilon_{mjk}B^j_aB^k_e\Psi_{ae}=(\hbox{det}B)(B^{-1})^d_i\psi_d\sim{0},
\end{eqnarray}
\noindent
where $\psi_d=\epsilon_{dae}\Psi_{ae}$ is the antisymmetric part of $\Psi_{ae}$.  This is equivalent to the diffeomorphism constraint $H_i$ owing to the nondegeneracy of $B^i_a$, and we will often use $H_i$ and $\psi_d$ interchangeably in this paper.  The equation of motion for the lapse function $N$, the analogue of the Hamilton equation for its conjugte momentum $\Pi_N$, is given by
\begin{eqnarray}
\label{OPTION76}
{{\delta{I}_{Inst}} \over {\delta{N}}}=(\hbox{det}B)^{1/2}\sqrt{\hbox{det}\Psi}\bigl(\Lambda+\hbox{tr}\Psi^{-1}\bigr)=0.
\end{eqnarray}
\noindent
Nondegeneracy of $\Psi_{ae}$ and the magnetic field $B^i_e$ implies that on-shell, the following relation must be satisfied
\begin{eqnarray}
\label{HAMM}
\Lambda+\hbox{tr}\Psi^{-1}=0,
\end{eqnarray}
\noindent
which we will similarly take as synonymous with the Hamiltonian constraint.  The equation of motion for $\Psi_{ae}$ is 
\begin{eqnarray}
\label{HODGE}
{{\delta{I}_{Inst}} \over {\delta\Psi_{ae}}}=B^k_eF^a_{0k}+\epsilon_{kjm}B^k_eB^j_aN^m+iN\sqrt{\hbox{det}B}\sqrt{\hbox{det}\Psi}(\Psi^{-1}\Psi^{-1})^{ea}\sim{0},
\end{eqnarray}
\noindent
up to a term proportional to (\ref{HAMM}) which we have set weakly equal to zero.  One could attempt to define a momentum conjugate to $\Psi_{ae}$, for which (\ref{HODGE}) would be the associated Hamilton's equation of motion.  But since $\Psi_{ae}$ forms part of the canonical structure of (\ref{OPTION78}), then our interpretation is that this is not technically correct.\footnote{This is because (\ref{HODGE}) contains a velocity $\dot{A}^a_k$ within $F^a_{0k}$ and will therefore be regarded as an evolution equation rather than a constraint.  This is in stark contrast with (\ref{SHIFT}) and (\ref{OPTION76}), which are genuine constraint equations due to the absence of any velocities.}\par
\indent  
The equation of motion for the connection $A^a_{\mu}$ is given by
\begin{eqnarray}
\label{MOTION11}
{{\delta{I}_{Inst}} \over {\delta{A}^a_{\mu}}}
\sim\epsilon^{\mu\sigma\nu\rho}D_{\sigma}(\Psi_{ae}F^e_{\nu\rho})
-{i \over 2}\delta^{\mu}_iD^{ij}_{da}\Bigl(4\epsilon_{mjk}N^mB^k_e\Psi_{[de]}\nonumber\\
+N(B^{-1})^d_j\sqrt{\hbox{det}B}\sqrt{\hbox{det}\Psi}\bigl(\Lambda+\hbox{tr}\Psi^{-1}\bigr)\Bigr),
\end{eqnarray}
\noindent
where we have defined
\begin{eqnarray}
\label{DECFIN}
\overline{D}^{ji}_{ea}(x,y)\equiv{\delta \over {\delta{A}^a_i(x)}}B^j_e(y)=\epsilon^{jki}\bigl(-\delta_{ae}\partial_k+f_{eda}A^d_k\bigr)\delta^{(3)}(x,y);~~\overline{D}^{0i}_{ea}\equiv{0}.
\end{eqnarray}
\noindent
The terms in large round brackets in (\ref{MOTION11}) vanish weakly, since they are proportional to the constraints (\ref{SHIFT}) and (\ref{HAMM}) and their spatial derivatives.  For the purposes of this paper we will 
regard (\ref{MOTION11}) as synonymous with
\begin{eqnarray}
\label{MOTION111}
\epsilon^{\mu\sigma\nu\rho}D_{\sigma}(\Psi_{ae}F^e_{\nu\rho})\sim{0}.
\end{eqnarray}
\noindent
In an abuse of notation, we will treat (\ref{HODGE}) and (\ref{MOTION111}) as strong equalities in this paper.  This will be justified once we have completed the demonstration that the constraint surface 
defined collectively by (\ref{SHIFT}), (\ref{OPTION76}) and the Gauss' constraint from (\ref{MOTION111}) is indeed preserved under time evolution.  As a note prior to proceeding we will often make the identification 
\begin{eqnarray}
\label{SHORTHAND}
N(\hbox{det}B)^{1/2}\sqrt{\hbox{det}\Psi}\equiv\sqrt{-g}
\end{eqnarray}
\noindent
as a shorthand notation, to avoid cluttering many of the derivations which follow in this paper.

\subsection{Internal consistency of the equations of motion}

Prior to embarking upon the issue of consistency of time evolution of the initial value constraints, we will check for internal consistency of $I_{Inst}$, which entails probing of the physical content implied by (\ref{MOTION111}) and (\ref{HODGE}).  First, equation (\ref{MOTION111}) can be decomposed into its spatial and temporal parts as
\begin{eqnarray}
\label{DECOMPOSE1}
D_i(\Psi_{bf}B^i_f)=0;~~D_0(\Psi_{bf}B^i_f)=\epsilon^{ijk}D_j(\Psi_{bf}F^f_{0k}).
\end{eqnarray}
\noindent
The first equation of (\ref{DECOMPOSE1}) is the Gauss' law constraint of a SO(3) Yang--Mills theory, when one makes the identification of $\Psi_{bf}B^i_f\sim{E}^i_b$ with the Yang--Mills electric field.  The Maxwell 
equations for $U(1)$ gauge theory with sources $(\rho,\vec{J})$, in units where $c=1$, are given by 
\begin{eqnarray}
\label{MAXWELL}
\vec{\nabla}\cdot\vec{B}=0;~~\dot{B}=-\vec{\nabla}\times\vec{E}=0;~~\vec{\nabla}\cdot\vec{E}=\rho;~~\dot{\vec{E}}=-\vec{J}+\vec{\nabla}\times\vec{B}.
\end{eqnarray}
\noindent
Equations (\ref{DECOMPOSE1}) can be seen as a generalization of the first two equations of (\ref{MAXWELL}) to $SO(3)$ nonabelian gauge theory in flat space when one: (i) identifies $F^f_{0k}\equiv{E}^f_k$ with the $SO(3)$ generalization of the electric field $\vec{E}$, and (ii) one chooses $\Psi_{ae}=k\delta_{ae}$ for some numerical constant $k$.\par
\indent  
When $\rho=0$ and $\vec{J}=0$, then one has the vacuum theory and equations (\ref{MAXWELL}) are invariant under the transformation
\begin{eqnarray}
\label{MAXWELL1}
(\vec{E},\vec{B})\longrightarrow(-\vec{B},\vec{E}).
\end{eqnarray}
\noindent
Then the second pair of equations of (\ref{MAXWELL}) become implied by the first pair.  This is the condition that the Abelian 
curvature $F_{\mu\nu}$, where $F_{0i}=E_i$ and $\epsilon_{ijk}F_{jk}=B_i$, is Hodge self-dual with respect to the metric of a conformally flat spacetime.  But equations (\ref{DECOMPOSE1}) for more general $\Psi_{ae}$ encode gravitational degrees of freedom, which as shown in \cite{EYOITA} generalizes the concept of self-duality to more general spacetimes solving the Einstein equations.  Let us first attempt to derive the analogue for (\ref{DECOMPOSE1}) of the second pair of (\ref{MAXWELL}) in the vacuum case.  Acting on the first equation of (\ref{DECOMPOSE1}) with $D_0$ yields
\begin{eqnarray}
\label{DECOMPOSE2}
D_0D_i(\Psi_{bf}B^i_f)=D_iD_0(\Psi_{bf}B^i_f)+[D_0,D_i](\Psi_{bf}B^i_f)=0.
\end{eqnarray}
\noindent
Substituting the second equation of (\ref{DECOMPOSE1}) into the first term on the right hand side of (\ref{DECOMPOSE2}) and using the definition of temporal curvature as the commutator of covariant derivatives on the second term we have
\begin{eqnarray}
\label{DECOMPOSE3}
D_i(\epsilon^{ijk}D_j(\Psi_{bf}F^f_{0k}))+f_{bcd}F^c_{0i}\Psi_{df}B^i_f
=f_{bcd}\bigl(B^k_cF^f_{0k}+B^k_fF^c_{0k}\bigr)\Psi_{df}=0
\end{eqnarray}
\noindent
where we have also used the spatial part of the commutator $\epsilon^{ijk}D_iD_jv_a=f_{abc}B^k_bv_c$.  Note that the term in brackets in (\ref{DECOMPOSE3}) is symmetric in $f$ and $c$, and also forms the symmetric part of the left hand side of (\ref{HODGE})
\begin{eqnarray}
\label{DECOMPOSE}
B^i_fF^b_{0i}+i\sqrt{-g}(\Psi^{-1}\Psi^{-1})^{fb}+\epsilon_{ijk}B^i_fB^j_bN^k=0,
\end{eqnarray}
\noindent
re-written here for completeness.  To make progress from (\ref{DECOMPOSE3}), we will substitute (\ref{DECOMPOSE}) into (\ref{DECOMPOSE3}).  This causes the last term of (\ref{DECOMPOSE}) to drop out due to antisymmetry, which leaves
 us with 
\begin{eqnarray}
\label{DECOMPOSE4}
-i\sqrt{-g}f_{bcd}\Bigl(\Psi_{df}(\Psi^{-1}\Psi^{-1})^{fc}+\Psi_{df}(\Psi^{-1}\Psi^{-1})^{fc}\Bigr)=-2i\sqrt{-g}f_{bcd}\Psi^{-1}_{dc}.
\end{eqnarray}
\noindent
The equations are consistent only if (\ref{DECOMPOSE4}) vanishes, which is the requirement that $\Psi_{ae}=\Psi_{ea}$ be symmetric.  This of course is the requirement that the diffeomorphism constraint (\ref{SHIFT}) be satisfied.  So the analogue of the second pair of (\ref{MAXWELL}) in the vacuum case must be encoded in the requirement that $\Psi_{ae}=\Psi_{ea}$ be symmetric.\par
\indent

\section{The time evolution equations}

We must now verify that the initial value constraints are preserved under time evolution defined by the equations of motion (\ref{HODGE}) and (\ref{MOTION11}).  These equations are respectively the Hodge duality condition 
\begin{eqnarray}
\label{BRINGIT}
B^k_fF^b_{0k}+i\sqrt{-g}(\Psi^{-1}\Psi^{-1})^{fb}+\epsilon_{ijk}N^iB^j_bB^k_f=0,
\end{eqnarray}
\noindent
and one of the Bianchi identity-like equations
\begin{eqnarray}
\label{BRINGIT1}
\epsilon^{ijk}D_j(\Psi_{ae}F^e_{ok})=D_0(\Psi_{ae}B^i_e).
\end{eqnarray}
\noindent
Since the initial value constraints were used to obtain the second line of (\ref{BRINGIT}) from (\ref{OPTION78}), then we must verify that these constraints are preserved under time evolution as a requirement of 
consistency.  Using $F^b_{0i}=\dot{A}^b_i-D_iA^b_0$ and defining
\begin{eqnarray}
\label{BRINGIT2}
\sqrt{-g}(B^{-1})^f_i(\Psi^{-1}\Psi^{-1})^{fb}+\epsilon_{mnk}N^mB^n_b\equiv{i}H^b_k,
\end{eqnarray}
\noindent
Then equation (\ref{BRINGIT}) can be written as a time evolution equation for the connection, which is not the same as a constraint equation as noted earlier
\begin{eqnarray}
\label{BRINGIT3}
F^b_{0i}=-iH^b_i\longrightarrow\dot{A}^b_i=D_iA^b_0-iH^b_i.
\end{eqnarray}
\noindent
From equation (\ref{BRINGIT3}) we can obtain the following equation governing time evolution equation for the magnetic field
\begin{eqnarray}
\label{MOOT2}
\dot{B}^i_e=\epsilon^{ijk}D_j\dot{A}^e_k=\epsilon^{ijk}D_j\bigl(D_kA^e_0-iH^e_k\bigr)\nonumber\\
=f_{ebc}B^i_bA^c_0-i\epsilon^{ijk}D_jH^e_k=-\delta_{\vec{\theta}}B^i_e-i\epsilon^{ijk}D_jH^e_k,
\end{eqnarray}
\noindent
which will be useful.  On the first term on the right hand side of (\ref{MOOT2}) we have used the definition of the curvature as the commutator of covariant derivatives.  The notation $\delta_{\vec{\theta}}$ in (\ref{MOOT2}) suggests that that $B^i_e$ transforms as a $SO(3,C)$ vector under gauge transformations parametrized by $\theta^b\equiv{A}^b_0$.\footnote{We will make the identification with $SO(3,C)$ gauge transformations later in this paper when we bring in the relation of $I_{Inst}$ with the Ashtekar variables.}  Since we have not specified anything about the canonical structure of $I_{Inst}$, then $\delta_{\vec{\theta}}$ as used 
in (\ref{MOOT2}) and in (\ref{MOOT4}) should at this stage simply be regarded as a definition useful for shorthand notation.\par
\indent
We will now apply the Liebnitz rule in conjunction with the definition of the temporal covariant derivatives to (\ref{BRINGIT1}) to determine the equation governing the time evolution of $\Psi_{ae}$.  This is given by
\begin{eqnarray}
\label{MOOT1}
D_0(\Psi_{ae}B^i_e)=B^i_e\dot{\Psi}_{ae}+\Psi_{ae}\dot{B}^i_e+f_{abc}A^b_0(\Psi_{ce}B^i_e)=\epsilon^{ijk}D_j(\Psi_{ae}F^e_{0k}).
\end{eqnarray}
\noindent
Substituting (\ref{MOOT2}) and (\ref{BRINGIT3}) into the left and right hand sides of (\ref{MOOT1}), we have
\begin{eqnarray}
\label{MOOT3}
B^i_e\dot{\Psi}_{ae}+\Psi_{ae}\bigl(f_{ebc}B^i_bA^c_0-i\epsilon^{ijk}D_jH^e_k\bigr)+f_{abc}A^b_0(\Psi_{ce}B^i_e)=-i\epsilon^{ijk}D_j(\Psi_{ae}H^e_k).
\end{eqnarray}
\noindent
In what follows, it will be convenient to use the following transformation properties for $\Psi_{ae}$ as $A^a_i$ under $SO(3,C)$ gauge transformations
\begin{eqnarray}
\label{MOOT4}
\delta_{\vec{\theta}}\Psi_{ae}=\bigl(f_{abc}\Psi_{ce}+f_{ebc}\Psi_{ac}\bigr)A^b_0;~~\delta_{\vec{\theta}}A^a_i=-D_iA^a_0;~~\delta_{\vec{\theta}}B^i_e=-f_{ebc}B^i_bA^c_0.
\end{eqnarray}
Then using (\ref{MOOT4}), the time evolution equations for the phase space variables $\Omega_{Inst}$ can be written in the following compact form
\begin{eqnarray}
\label{MOOT5}
\dot{A}^b_i=-\delta_{\vec{\theta}}A^b_i-iH^b_i;~~
\dot{\Psi}_{ae}=-\delta_{\vec{\theta}}\Psi_{ae}-i\epsilon^{ijk}(B^{-1})^e_i(D_j\Psi_{af})H^f_k.
\end{eqnarray}
\noindent
We have found evolution equations for $\Psi_{ae}$ and $A^a_i$ from the covariant equations of $A^a_{\mu}$ and the Hodge-duality condition  We have obtained these without using Poisson brackets, and by assuming that the Hamiltonian and diffeomorphism constraints are satisfied.  Therefore the first order of business is then to check for the preservation of the initial value constraints under the time evolution generated by (\ref{MOOT5}).  This means that we must check that the time evolution of the diffeomorphism, Gauss' law and Hamiltonian constraints are combinations of terms proportional to the same constraints and their spatial derivatives, and terms which vanish when the constraints hold.\footnote{This includes any nonlinear function of linear order or higher in the constraints, a situation which involves the diffeomorphism constraint.}  These constraints are given by
\begin{eqnarray}
\label{THESECONSTRAINTS}
\textbf{w}_e\{\Psi_{ae}\}=0;~~(\hbox{det}B)(B^{-1})^d_i\psi_d=0;~~(\hbox{det}B)^{1/2}\sqrt{\hbox{det}\Psi}\bigl(\Lambda+\hbox{tr}\Psi^{-1}\bigr)=0
\end{eqnarray}
\noindent
where $(\hbox{det}B)\neq{0}$ and $(\hbox{det}\Psi)\neq{0}$.  We will occasionally make the identification
\begin{eqnarray}
\label{IDENTI}
N(\hbox{det}B)^{1/2}(\hbox{det}\Psi)^{1/2}\equiv\sqrt{-g}
\end{eqnarray}
\noindent
for a shorthand notation.  Additionally, the following definitions are provided for the vector fields appearing in the Gauss' constraint
\begin{eqnarray}
\label{IDENTI1}
\textbf{w}_e=B^i_eD_i;~~\textbf{v}_e=B^i_e\partial_i
\end{eqnarray}
\noindent
where $D_i$ is the $SO(3,C)$ covariant derivative with respect to the connection $A^a_i$.  Equations (\ref{THESECONSTRAINTS}) are the equations of motion for the auxilliary fields $A^a_0$, $N^i$ and $N$.

\section{Consistency of the diffeomorphism constraint under time evolution}
The diffeomorphism constraint is directly proportional to $\psi_d=\epsilon_{dae}\Psi_{ae}$, the antisymmetric part of $\Psi_{ae}$.  So to establish the consistency condition for this constraint, it suffices to show that the antisymmetric part of the second equation of (\ref{MOOT5}) weakly vanishes.  This is given by
\begin{eqnarray}
\label{DEAF}
\epsilon_{dae}\dot{\Psi}_{ae}=-\delta_{\vec{\theta}}(\epsilon_{dae}\Psi_{ae})-i\epsilon_{dae}\epsilon^{ijk}(B^{-1})^e_i(D_j\Psi_{af})H^f_k,
\end{eqnarray}
\noindent
which splits into two terms.  Using (\ref{MOOT4}), one finds that the first term of (\ref{DEAF}) is given by
\begin{eqnarray}
\label{DEEF}
-\epsilon_{dae}\delta_{\vec{\theta}}\Psi_{ae}=-\epsilon_{dae}\bigl(f_{abc}\Psi_{ce}+\Psi_{ac}f_{ebc}\bigr)A^b_0\nonumber\\
=-\Bigl(\bigl(\delta_{eb}\delta_{dc}-\delta_{ec}\delta_{bd}\bigr)\Psi_{ce}+\bigl(\delta_{db}\delta_{ac}-\delta_{dc}\delta_{ab}\bigr)\Psi_{ac}\Bigr)A^b_0\nonumber\\
=-\bigl(\Psi_{db}-\delta_{bd}\hbox{tr}\Psi+\delta_{db}\hbox{tr}\Psi-\Psi_{bd}\bigr)A^b_0=2\Psi_{[bd]}A^b_0=-\epsilon_{dbh}A^b_0\psi_h,
\end{eqnarray}
\noindent
which is proportional to the diffeomorphism constraint.  The second term of (\ref{DEEF}) has two contributions due to $H^f_k$ as defined in (\ref{BRINGIT2}).  The first contribution reduces to
\begin{eqnarray}
\label{DEEFONE}
-i\epsilon_{dae}\epsilon^{ijk}(B^{-1})^e_i(D_j\Psi_{af})(H_{(1)})^f_k\nonumber\\
=-i\epsilon_{dae}\epsilon^{ijk}(B^{-1})^e_i(D_j\Psi_{af})\sqrt{-g}(B^{-1})^g_k(\Psi^{-1}\Psi^{-1})^{gf}\nonumber\\
=i\epsilon_{dae}(\hbox{det}B)^{-1}\epsilon^{egh}(\Psi^{-1}\Psi^{-1})^{gf}B_h^jD_j\Psi_{af}\nonumber\\
=i(\hbox{det}B)^{-1}(\Psi^{-1}\Psi^{-1})^{gf}\bigl(\delta^g_d\delta^h_a-\delta^g_a\delta^h_d\bigr)\textbf{v}_v\{\Psi_{af}\}\nonumber\\
=i(\hbox{det}B)^{-1}(\Psi^{-1}\Psi^{-1})^{gf}\bigl(\delta^g_d\textbf{v}_a\{\Psi_{af}\}-\textbf{v}_d\{\Psi_{gf}\}\bigr)\nonumber\\
=i(\hbox{det}B)^{-1}\Bigl[(\Psi^{-1}\Psi^{-1})^{df}G_f+\textbf{v}_d\{\Lambda+\hbox{tr}\Psi^{-1}\}\Bigr].
\end{eqnarray}
\noindent
The first term on the final right hand side of (\ref{DEEFONE}) is the Gauss' constraint and the second term is the derivative of a term direction proportional to the Hamiltonian constraint.\footnote{We have added in a term $\Lambda$, which can be regarded as a constant of integration with respect to the spatial derivatives from $\textbf{v}_d$.}  The second contribution to the second term of (\ref{DEAF}) is given by
\begin{eqnarray}
\label{DEEFTWO}
\epsilon_{dae}\epsilon^{ijk}(B^{-1})^e_i(D_j\Psi_{af})(H_{(2)})^f_k
=\epsilon_{dae}\epsilon^{ijk}(B^{-1})^e_i(D_j\Psi_{af})\epsilon_{mnk}N^mB^n_f\nonumber\\
=\epsilon_{dac}\bigl(\delta^i_m\delta^j_n-\delta^i_n\delta^j_m\bigr)(B^{-1})^e_i(D_j\Psi_{af})N^mB^n_f\nonumber\\
=\epsilon_{dae}N^i(B^{-1})^e_i\textbf{v}_f\{\Psi_{af}\}-N^jD_j(\epsilon_{dae}\Psi_{ae})
=\epsilon_{dae}N^i(B^{-1})^e_fG_a-N^jD_j\psi_d.
\end{eqnarray}
\noindent
The result is that the time evolution of the diffeomorphism constraint is directly proportional to
\begin{eqnarray}
\label{DEEFTHREE}
\dot{\psi}_d=\Bigl[i(\hbox{det}B)^{-1}(\Psi^{-1}\Psi^{-1})^{da}+\epsilon_{dae}N^i(B^{-1})^e_i\Bigr]G_a\nonumber\\
+\bigl(A^b_0\epsilon_{bdh}-\delta_{dh}N^jD_j\bigr)\psi_h+i(\hbox{det}B)^{-1}\textbf{v}_d\{(-g)^{-1/2}H\},
\end{eqnarray}
\noindent
which is a linear combination of terms proportional to the constraints (\ref{THESECONSTRAINTS}) and their spatial derivatives.  The result is that the diffeomorphism constraint $H_i=0$ is consistent with respect to the Hamiltonian evolution generated by the equations (\ref{MOOT5}).  So it remains to verify consistency of Gauss' law and the Hamiltonian constraints $G_a$ and $H$.  

\section{Consistency of the Gauss' constraint under time evolution}
Having verified the consistency of the diffeomorphism constraint under time evolution, we now move on to the Gauss' constraint.  Application of the Liebnitz rule to the first equation of (\ref{THESECONSTRAINTS}) yields
\begin{eqnarray}
\label{MOOT6}
\dot{G}_a=\dot{B}^i_eD_i\Psi_{ae}+B^i_eD_i\dot{\Psi}_{ae}+B^i_e\bigl(f_{abf}\Psi_{fe}+f_{ebg}\Psi_{ag}\bigr)\dot{A}^a_i.
\end{eqnarray}
\noindent
Upon substituion of (\ref{MOOT2}) and (\ref{MOOT5}) into (\ref{MOOT6}), we have
\begin{eqnarray}
\label{MOOT7}
\dot{G}_a=\bigl(-\delta_{\vec{\theta}}B^i_e-i\epsilon^{ijk}D_jH^e_k\bigr)D_i\Psi_{ae}+B^m_eD_m\bigl(-\delta_{\vec{\theta}}\Psi_{ae}-i\epsilon^{ijk}(B^{-1})^e_i(D_j\Psi_{af})H^f_k\bigr)\nonumber\\
+B^i_e\bigl(f_{abf}\Psi_{fe}+f_{ebg}\Psi_{ag}\bigr)\bigl(-\delta_{\vec{\theta}}A^b_i-iH^b_i\bigr).
\end{eqnarray}
\noindent
Using the Liebniz rule to combine the $\delta_{\vec{\theta}}$ terms of (\ref{MOOT7}), we have
\begin{eqnarray}
\label{MOOT8}
\dot{G}_a=-\delta_{\vec{\theta}}G_a-i\epsilon^{ijk}\Bigl[(D_jH^e_k)D_i\Psi_{ae}+B^m_eD_m((B^{-1})^e_i(D_j\Psi_{af})H^f_k)\Bigr]\nonumber\\
-i\bigl(f_{abf}\Psi_{fe}+f_{ebg}\Psi_{ag}\bigr)B^i_eH^b_i.
\end{eqnarray}
\noindent
The requirement of consistency is that we must show that the right hand side of (\ref{MOOT8}) vanishes weakly.  First, we will show that the third term on the right hand side of (\ref{MOOT8}) vanishes up to terms of linear order and higher in the diffeomorphism constraint.  This term, up to an insignificant numerical factor, has two contributions.  The first contribution is
\begin{eqnarray}
\label{MOOT9}
\bigl(f_{abf}\Psi_{fe}+f_{ebg}\Psi_{ag}\bigr)B^i_e(H_{(1)})^b_i=\sqrt{-g}\bigl(f_{abf}\Psi_{fe}+f_{ebg}\Psi_{ag}\bigr)(\Psi^{-1}\Psi^{-1})^{eb}\nonumber\\
=\sqrt{-g}\bigl(f_{abf}(\Psi^{-1})^{fb}+f_{ebg}(\Psi^{-1}\Psi^{-1})^{eb}\Psi_{ag}\bigr)\sim\delta^{(1)}_a(\vec{\psi})\sim{0},
\end{eqnarray}
\noindent
which is directly proportional to a nonlinear function of first order in $\psi_d$ which is proportional to the diffeomorphism constraint.  The second contribution to the third term on the right hand side of (\ref{MOOT8}) is
\begin{eqnarray}
\label{MOOT10}
\bigl(f_{abf}\Psi_{fe}+f_{ebg}\Psi_{ag}\bigr)B^i_e(H_{(2)})^b_i=\bigl(f_{abf}\Psi_{fe}+f_{ebg}\Psi_{ag}\bigr)\epsilon_{kmn}N^kB^m_eB^n_b\nonumber\\
=\bigl(f_{abf}\Psi_{fe}+f_{ebg}\Psi_{ag}\bigr)(\hbox{det}B)N^k(B^{-1})^d_k\epsilon_{deb}\nonumber\\
=(\hbox{det}B)N^k(B^{-1})^d_k\bigl(\bigl(\delta_{fd}\delta_{ae}-\delta_{fe}\delta_{ad}\bigr)\Psi_{fe}+2\delta_{dg}\Psi_{ag}\bigr)\nonumber\\
=(\hbox{det}B)N^k(B^{-1})^d_k\bigl(\Psi_{da}-\delta_{ad}\hbox{tr}\Psi+2\Psi_{ad}\bigr)\equiv\delta^{(2)}_a(\vec{N})
\end{eqnarray}
\noindent
which does not vanish, and neither is it expressible as a constraint.  For the Gauss' law constraint to be consistent under time evolution, a necessary condition is that this $\delta^{(2)}_a(\vec{N})$ term must be exactly cancelled by another term arising from the variation.\par
\indent
Let us expand the terms in square brackets in (\ref{MOOT8}).  This is given, using the Liebniz rule on the second term, by
\begin{eqnarray}
\label{MOOT11}
\epsilon^{ijk}(D_jH^e_k)(D_i\Psi_{ae})+\epsilon^{ijk}B^m_eD_m((B^{-1})^e_i(D_j\Psi_{ae})H^f_k)\nonumber\\
=\epsilon^{ijk}(D_jH^e_k)(D_i\Psi_{ae})-\epsilon^{ijk}B^m_e(B^{-1})^e_n(D_mB^n_g)(B^{-1})^g_i(D_j\Psi_{af})H^f_k\nonumber\\
+\epsilon^{mjk}(D_mD_j\Psi_{af})H^f_k+\epsilon^{mjk}(D_j\Psi_{af})(D_mH^f_k).
\end{eqnarray}
\noindent
The first and last terms on the right hand side of (\ref{MOOT11}) cancel, which can be seen by relabelling of indices.  Upon application of the definition of curvature as the commutator of covariant derivatives to the third term, then (\ref{MOOT11}) reduces to
\begin{eqnarray}
\label{MOOT12}
-\epsilon^{ijk}(D_nB^n_g)(B^{-1})^g_i(D_j\Psi_{af})H^f_k+H^f_kB^k_b\bigl(f_{abc}\Psi_{cf}+f_{fbc}\Psi_{ac}\bigr).
\end{eqnarray}
\noindent
The first term of (\ref{MOOT12}) vanishes on account of the Bianchi identity and the second term contains two contributions which we must evaluate.  The first contribution is given by
\begin{eqnarray}
\label{MOOTY}
(H_{(2)})^f_kB^k_a\bigl(f_{abc}\Psi_{cf}+f_{fbc}\Psi_{ac}\bigr)=(\hbox{det}B)N^k(B^{-1})^d_k\epsilon_{dbf}\bigl(f_{abc}\Psi_{cf}+f_{fbc}\Psi_{ac}\bigr)\nonumber\\
=(\hbox{det}B)N^k(B^{-1})^d_k\bigl(\bigl(\delta_{da}\delta_{fc}-\delta_{dc}\delta_{fa}\bigr)\Psi_{cf}-2\delta_{dc}\Psi_{ac}\bigr)\nonumber\\
=(\hbox{det}B)N^k(B^{-1})^d_k\bigl(\delta_{da}\hbox{tr}\Psi-\Psi_{da}-2\Psi_{ad}\bigr)=-\delta_a^{(2)}(\vec{N}),
\end{eqnarray}
\noindent
with $\delta_a^{(2)}(\vec{N})$ as given in (\ref{MOOT9}).  So putting the results of (\ref{MOOT11}), (\ref{MOOT12}) and (\ref{MOOTY}) into (\ref{MOOT8}), we have
\begin{eqnarray}
\label{MOOTA}
\dot{G}_a=-\delta_{\vec{\theta}}G_a+\delta^{(2)}_a(\vec{N})+\delta^{(1)}_a(\vec{\psi})+\delta^{(1)}_a(\vec{\psi})-\delta^{(2)}_a(\vec{N})
=-\delta_{\vec{\theta}}G_a+2\delta^{(1)}(\vec{\psi}).
\end{eqnarray}
\noindent
The velocity of the Gauss' law constraint is a linear combination of the Gauss' constraint with terms of the diffeomorphism constraint of linear order and higher.  Hence the time evolution of the Gauss' law constraint is consistent in the sense that we have defined, since $\delta^{(1)}(\vec{\psi})$ vanishes for $\psi_d=0$.

\section{Consistency of the Hamiltonian constraint under time evolution}

The time derivative of the Hamiltonian constraint, the third equation of (\ref{THESECONSTRAINTS}), is given by
\begin{eqnarray}
\label{GREENTUB}
\dot{H}=\bigl[{d \over {dt}}((\hbox{det}B)^{1/2}(\hbox{det}\Psi)^{1/2}\bigl](\Lambda+\hbox{tr}\Psi^{-1})+{{\sqrt{-g}} \over N}{d \over {dt}}(\Lambda+\hbox{tr}\Psi^{-1})
\end{eqnarray}
\noindent
which has split up into two terms.  The first term is directly proportional to the Hamiltonian constraint, therefore it is already consistent.  We will nevertheless expand it using (\ref{MOOT2}) and (\ref{MOOT5})
\begin{eqnarray}
\label{GREEN}
{1 \over 2}\bigl((B^{-1})^d_i\dot{B}^i_d+(\Psi^{-1})^{ae}\dot{\Psi}_{ae}\bigr)(\hbox{det}B)^{1/2}(\hbox{det}\Psi)^{1/2}(\Lambda+\hbox{tr}\Psi^{-1})\nonumber\\
={1 \over 2}\Bigl((B^{-1})^d_i\bigl(-\delta_{\vec{\theta}}B^i_d-i\epsilon^{ijk}D_jH^d_k\bigr)\nonumber\\
+(\Psi^{-1})^{ae}\bigl(-\delta_{\vec{\theta}}\Psi_{ae}-i\epsilon^{ijk}(B^{-1})^e_i(D_j\Psi_{af})H^f_k\Bigr)H.
\end{eqnarray}
\noindent
We will be content to compute the $\delta_{\vec{\theta}}$ terms of (\ref{GREEN}).  These are
\begin{eqnarray}
\label{GREEN1}
(B^{-1})^d_i\delta_{\vec{\theta}}B^i_d=(B^{-1})^d_if_{dbf}B^i_bA^f_0=\delta_{db}f_{dbf}A^f_0=0
\end{eqnarray}
\noindent
on account of antisymmetry of the structure constants, and
\begin{eqnarray}
\label{GREEN2}
(\Psi^{-1})^{ea}\delta_{\vec{\theta}}\Psi_{ae}=(\Psi^{-1})^{ea}\bigl(f_{abf}\Psi_{fe}+f_{ebg}\Psi_{ag}\bigr)=0,
\end{eqnarray}
\noindent
also due to antisymmetry of the structure constants.  We have shown that the first term on the right hand side of (\ref{GREENTUB}) is consistent with respect to time evolution.  To verify consistency of the Hamiltonian constraint under time evolution, it remains to show that the second term is weakly equal to zero.  It suffices to show this just for the second term, in brackets, of (\ref{GREENTUB})
\begin{eqnarray}
\label{MOOT17}
{d \over {dt}}(\Lambda+\hbox{tr}\Psi^{-1})=-(\Psi^{-1}\Psi^{-1})^{fe}\dot{\Psi}_{ef}\nonumber\\
=(\Psi^{-1}\Psi^{-1})^{ef}\bigl(\delta_{\vec{\theta}}\Psi_{ae}-i\epsilon^{ijk}(B^{-1})^e_i(D_j\Psi_{af})H^f_k\bigr),
\end{eqnarray}
\noindent
where we have used (\ref{MOOT5}).  Equation (\ref{MOOT17}) has split up into two terms, of which the first term is
\begin{eqnarray}
\label{MOOT18}
(\Psi^{-1}\Psi^{-1})^{ea}\delta_{\vec{\theta}}\Psi_{ae}=(\Psi^{-1}\Psi^{-1})^{ea}\bigl(f_{Abf}\Psi_{fe}+f_{ebg}\Psi_{ag}\bigr)A^b_0\nonumber\\
\bigl(f_{abf}(\Psi^{-1})^{fa}+f_{ebg}(\Psi^{-1})^{eg}\bigr)A^b_0=m(\vec{\psi})\sim{0}
\end{eqnarray}
\noindent
which vanishes weakly since it is a nonlinear function of at least linear order in $\psi_d$.  The second term of (\ref{MOOT17}) splits into two terms which we must evaluate.  The first contribution is proportional to
\begin{eqnarray}
\label{MOOT19}
(\Psi^{-1}\Psi^{-1})^{ea}\epsilon^{ijk}(B^{-1})^e_i(D_j\Psi_{af})(H_{(1)})^f_k\nonumber\\
=\sqrt{-g}(\Psi^{-1}\Psi^{-1})^{ea}\epsilon^{ijk}(B^{-1})^e_i(D_j\Psi_{af})(B^{-1})^d_k(\Psi^{-1}\Psi^{-1})^{df}\nonumber\\
=-\sqrt{-g}(\Psi^{-1}\Psi^{-1})^{ea}(\Psi^{-1}\Psi^{-1})^{df}(\hbox{det}B)^{-1}\epsilon^{edg}B^j_gD_j\Psi_{af}\nonumber\\
=-\sqrt{-g}(\hbox{det}B)^{-1}\epsilon^{edg}(\Psi^{-1}\Psi^{-1})^{ea}(\Psi^{-1}\Psi^{-1})^{df}\textbf{v}_g\{\Psi_{af}\}\equiv\textbf{v}\{\vec{\psi}\}
\end{eqnarray}
\noindent
for some vector field $\textbf{v}$.  We have used the fact that the term in (\ref{MOOT19}) quartic in $\Psi^{-1}$ in antisymmetric in $a$ and $f$ due to the epsilon symbol.  Hence $\Psi_{af}$ as acted upon by $\textbf{v}_g$ can only appear in an antisymmetric combination, and is therefore proportional to the diffeomorphism constraint $\psi_d$ whose spatial derivatives weakly vanish.  Hence (\ref{MOOT19}) presents a consistent contribution to the time evolution of $H$, which leaves remaining the second contribution to the second term of (\ref{MOOT17}).  This term is proportional to
\begin{eqnarray}
\label{MOOT20}
(\Psi^{-1}\Psi^{-1})^{ea}\epsilon^{ijk}(B^{-1})^e_i(D_j\Psi_{af})(H_{(2)})^f_k\nonumber\\
=(\Psi^{-1}\Psi^{-1})^{ea}\epsilon^{ijk}(B^{-1})^e_i(D_j\Psi_{af})\epsilon_{mnk}N^mB^n_f\nonumber\\
=\bigl(\delta^i_m\delta^j_n-\delta^i_n\delta^j_m\bigr)(B^{-1})^e_iB^n_f(\Psi^{-1}\Psi^{-1})^{ea}(D_j\Psi_{af})\nonumber\\
=\bigl(N^i(B^{-1})^e_iB^j_f-\delta_{ef}N^j\bigr)(\Psi^{-1}\Psi^{-1})^{ea}(D_j\Psi_{af})\nonumber\\
=(-g)^{-1/2}N^iH^a_i\textbf{v}_f\{\Psi_{af}\}-(\Psi^{-1}\Psi^{-1})^{fa}(N^jD_j\Psi_{af})\nonumber\\
=(-1)^{-1/2}N^iH^a_iG_a-N^jD_j(\Lambda+\hbox{tr}\Psi^{-1}).
\end{eqnarray}
\noindent
The first term on the final right hand side of (\ref{MOOT20}) is proportional to the Gauss' law constraint, and the second term is proportional to the derivative of the Hamiltonian constraint.  To obtain this second term we have added in $\Lambda$ as a constant of differentiation with respect to $\partial_j$.  Substituting (\ref{MOOT18}), (\ref{MOOT19}) and (\ref{MOOT20}) into (\ref{MOOT17}), then we have
\begin{eqnarray}
\label{MUTATE}
\dot{H}=\sim\hat{O}(\vec{\psi})+(-g)^{-1/2}N^iH^a_iG_a+\hat{T}((-g)^{-1/2}H),
\end{eqnarray}
\noindent
where $\hat{O}$ and $\hat{T}$ are operators consisting of spatial derivatives acting to the right and $c$ numbers.  The time derivative of the Hamiltonian constraint is a linear combination of the Gauss' law and Hamiltonian constraints and its spatial derivatives, plus terms of linear order and higher in the diffeomorphism constraint and its spatial derivatives.  Hence the Hamiltonian constraint is consistent under time evolution.

\section{Recapitulation}
The final equations governing the time evolution of the initial value constraints are given weakly by
\begin{eqnarray}
\label{BOOTLEG}
\dot{\psi}_d=\Bigl[i(\hbox{det}B)^{-1}(\Psi^{-1}\Psi^{-1})^{da}+\epsilon_{dae}N^i(B^{-1})^e_i\Bigr]G_a\nonumber\\
+\bigl(A^b_0\epsilon_{bdh}-\delta_{dh}N^jD_j\bigr)\psi_h+i(\hbox{det}B)^{-1}\textbf{v}_d\{\Lambda+\hbox{tr}\Psi^{-1}\};\nonumber\\
\dot{G}_a=-f_{abc}A^b_0G_c+\delta^{(1)}_a(\vec{\psi});\nonumber\\
\dot{H}=\Bigl[-{i \over 2}\epsilon^{ijk}(B^{-1})^d_i(D_jH^d_k)+\epsilon^{ijk}(B^{-1})^e_i(\Psi^{-1})^{ae}(D_j\Psi_{af})H^f_k-N^j\partial_j\Bigr](\Lambda+\hbox{tr}\Psi^{-1})\nonumber\\
+(-g)^{-1/2}N^iH^a_iG_a
-\sqrt{-g}(\hbox{det}B)^{-1}\epsilon^{edg}(\Psi^{-2}\Psi^{-1})^{ea}(\Psi^{-1}\Psi^{-1})^{df}\textbf{v}_g\{\epsilon_{afh}\psi_h\}+m(\vec{\psi}).
\end{eqnarray}
\noindent
Equations (\ref{BOOTLEG}) show that all constraints derivable from the the action (\ref{OPTION78}) are preserved under time evolution, since their time derivatives yield linear combinations of the same set of constraints and their spatial derivatives.  There are no additional constraints generated which implies that the action (\ref{OPTION78}) is consistent in the Dirac sense.  On the other hand, we have not defined the canonical structure of (\ref{BOOTLEG}) or any Poisson brackets.\par
\indent
Equations (\ref{BOOTLEG}) can be written schematically in the following form
\begin{eqnarray}
\label{BOOTLEG1}
\dot{\vec{H}}\sim\vec{H}+\vec{G}+H;~~\dot{\vec{G}}\sim\vec{G}+\Phi(\vec{H});~~
\dot{H}\sim{H}+\vec{G}+\Phi(\vec{H}),
\end{eqnarray}
\noindent
where $\Phi$ is some nonlinear function of the diffeomorphism constraint $\vec{H}$, which is of at least first order in $\vec{H}$.  In the Hamiltonian formulation of a theory, one identifies time derivatives of a variable $f$ with via $\dot{f}=\{f,\boldsymbol{H}\}$ the Poisson brackets of the variable with the Hamiltonian $\boldsymbol{H}$.  So while we have not specified Poisson brackets, equation (\ref{BOOTLEG1}) implies the existence of Poisson brackets associated to some Hamiltonian $\boldsymbol{H}_{Inst}$ for the action (\ref{OPTION78}), with
\begin{eqnarray}
\label{BOOTLEG2}
\{\vec{H},\boldsymbol{H}_{Inst}\}\sim\vec{H}+\vec{G}+H;~~
\{\vec{G},\boldsymbol{H}_{Inst}\}\sim\vec{G}+\Phi(\vec{H});\nonumber\\
\{H,\boldsymbol{H}_{Inst}\}\sim{H}+\Phi(\vec{H})+\vec{G}.
\end{eqnarray}
\noindent
So the main result of this paper has been to demonstrate that the instanton representation of Plebanski gravity forms a consistent system, in the sense that the constraint surface is preserved under time evolution.  As a direction of future research we will compute the algebra of constraints for (\ref{OPTION78}) directly from its canonical structure.  Nevertheless it will be useful for the present paper to think of equations (\ref{BOOTLEG}) in the Dirac context, mainly for comparison with other formulations of general relativity.  This will bring us to the Ashtekar variables.

\section{Discussion: Relation of the instanton representation to the Ashtekar variables}
We will now provide the rationale for not following the Dirac procedure for constrained systems \cite{DIR} with respect to (\ref{OPTION78}), by comparison with the Ashekar formulation of GR.  The action for the instanton representation (\ref{OPTION78}) can be written in the following 3+1 decomposed form
\begin{eqnarray}
\label{CHEN88}
I_{Inst}=\int{dt}\int_{\Sigma}d^3x\Bigl[\Psi_{ae}B^i_e\dot{A}^a_i+A^a_0\textbf{w}_e\{\Psi_{ae}\}-\epsilon_{ijk}N^iB^j_aB^k_e\Psi_{ae}\nonumber\\
-iN(\hbox{det}B)^{1/2}(\hbox{det}\Psi)^{1/2}\bigl(\Lambda+\hbox{tr}\Psi^{-1}\bigr)\Bigr],
\end{eqnarray}
\noindent
which regards $\Psi_{ae}$ and $A^a_i$ as phase space variables.  But the phase space of (\ref{CHEN88}) is noncanonical since its symplectic two form
\begin{eqnarray}
\label{NONCANON}
\delta\boldsymbol{\theta}_{Inst}=\delta\Bigl(\int_{\Sigma}d^3x\Psi_{ae}B^i_e\delta{A}^a_i\Bigr)\nonumber\\
=\int_{\Sigma}d^3xB^i_e{\delta\Psi_{ae}}\wedge{\delta{A}^a_i}+\int_{\Sigma}d^3x\Psi_{ae}\epsilon^{ijk}{D_j(\delta{A}^e_k)}\wedge{\delta{A}^a_i},
\end{eqnarray}
\noindent
is not closed owing to the presence of the second term on the right hand side.  The initial stages of the Dirac procedure applied to (\ref{CHEN88}) state that the momentum conjugate to $A^a_i$ yields the primary constraint
\begin{eqnarray}
\label{CHENTIBLE}
\Pi^i_a={{\delta{I}_{Inst}} \over {\delta\dot{A}^a_i}}=\Psi_{ae}B^i_e.
\end{eqnarray}
\noindent
Then making the identification $\widetilde{\sigma}^i_a=\Pi^i_a$ and upon substitution into (\ref{CHENTIBLE}) and into (\ref{CHEN88}), one obtains the action
\begin{eqnarray}
\label{ASHTEK}
I_{Ash}=\int{dt}\int_{\Sigma}d^3x\Bigl[\widetilde{\sigma}^i_a\dot{A}^a_i+A^a_0G_a-N^iH_i-{i \over 2}\underline{N}H\Bigr],
\end{eqnarray}
\noindent
which is the action for the Ashtekar complex formalism of general relativity \cite{ASH2}, \cite{ASH3}, with $\widetilde{\sigma}^i_a$ being the densitized triad.  This is a totally constrained system with $(A^a_0,N^i,\underline{N})$, respectively the $SO(3,C)$ rotation angle $A^a_0$, the shift vector $N^i$ and the densitized lapse function $\underline{N}=N(\hbox{det}\widetilde{\sigma})^{-1/2}$ as auxilliary fields.  The constraints in (\ref{ASHTEK}) smearing the auxilliary fields are the Gauss' law, vector and Hamiltonian constraints
\begin{eqnarray}
\label{ACTIONASH1}
G_a=D_i\widetilde{\sigma}^i_a;~~H_i=\epsilon_{ijk}\widetilde{\sigma}^j_aB^k_a;~~H=\epsilon_{ijk}\epsilon^{abc}\widetilde{\sigma}^i_a\widetilde{\sigma}^j_b\Bigl({\Lambda \over 3}\widetilde{\sigma}^k_c+B^k_c\Bigr).
\end{eqnarray}
\noindent
From (\ref{ASHTEK}) one reads off the symplectic two form $\boldsymbol{\Omega}_{Ash}$ given by
\begin{eqnarray}
\label{SYMPL}
\boldsymbol{\Omega}_{Ash}=\int_{\Sigma}d^3x{\delta\widetilde{\sigma}^i_a}\wedge{\delta{A}^a_i}=\delta\Bigl(\int_{\Sigma}d^3x\widetilde{\sigma}^i_a\delta{A}^a_i\Bigr)=\delta\boldsymbol{\theta}_{Ash},
\end{eqnarray}
\noindent
which is the exact functional variation of the canonical one form $\boldsymbol{\theta}_{Ash}$.\par
\indent 
The actions (\ref{CHEN88}) and (\ref{ASHTEK}) are transformable into each other only under the condition $(\hbox{det}B)\neq{0}$ and $(\hbox{det}\Psi)\neq{0}$.  In (\ref{ASHTEK}) it is clear that $\widetilde{\sigma}^i_a$ and $A^a_i$ form a canonically conjugate pair, which suggests that (\ref{CHEN88}) is a noncanonical version of (\ref{ASHTEK}).  The constraints algebra for (\ref{ACTIONASH1}) is 
\begin{eqnarray}
\label{ALGEBRA12}
\{\vec{H}[\vec{N}],\vec{H}[\vec{M}]\}=H_k\bigl[N^{i}\partial^{k}M_i-M^{i}\partial^{k}N_i\bigr];\nonumber\\
\{\vec{H}[N],G_a[\theta^a]\}=G_a[N^{i}\partial_{i}\theta^a];\nonumber\\
\{G_a[\theta^a],G_b[\lambda^b]\}=G_{a}\bigl[f^a_{bc}\theta^{b}\lambda^c\bigr];\nonumber\\
\{H(\underline{N}),\vec{H}[\vec{N}]\}=H[N^{i}\partial_{i}\underline{N}\bigr]\nonumber\\
\{H(\underline{N}),G_a(\theta^a)\}=0;\nonumber\\
\bigl[H(\underline{N}),H(\underline{M})\bigr]
=H_{i}[\bigl(\underline{N}\partial_{j}\underline{M}
-\underline{M}\partial_{j}\underline{N}\bigr)H^{ij}],
\end{eqnarray}
\noindent
which is first class due to closure of the algebra, and is therefore consistent in the Dirac sense.  Let us consider (\ref{ALGEBRA12}) for each constraint with the total 
Hamiltonian $\boldsymbol{H}_{Ash}$ and compare with (\ref{BOOTLEG2}).  This is given schematically by
\begin{eqnarray}
\label{ALGEBRA13}
\{\vec{H},\boldsymbol{H}_{Ash}\}\sim\vec{H}+\vec{G}+H;~~
\{\vec{G},\boldsymbol{H}_{Ash}\}\sim\vec{G}+\vec{H};\nonumber\\
\{H,\boldsymbol{H}_{Ash}\}\sim{H}+\vec{H}.
\end{eqnarray}
Comparison of (\ref{ALGEBRA13}) with (\ref{BOOTLEG2}) shows an essentially similar structure for the top two lines involving $\vec{H}$ and $\vec{G}$.\footnote{The linearly versus nonlinearly of the diffeomorphism constraints on the right hand side is just a minor difference.}  But there is a marked dissimilarity with respect to the Hamiltonian constraint $H$.  Note that there is a Gauss' law constraint appearing in the right hand side of the last line of 
(\ref{BOOTLEG2}) whereas there is no such constraint on the corresponding right hand side of (\ref{ALGEBRA13}).  This means that while the Hamiltonian constraint is gauge-invariant under $SO(3,C)$ gauge-transformations as 
implied by (\ref{ALGEBRA12}) and (\ref{ALGEBRA13}), this is not the case in (\ref{BOOTLEG2}).  This means that the action (\ref{OPTION78}), which as shown in \cite{EYOITA} describes general relativity for Petrov Types I, D and O, has a different role for the Gauss' law and Hamiltonian constraints than the action (\ref{ASHTEK}), which also describes general relativity.  Therefore $I_{Inst}$ and $I_{Ash}$ at some level correspond to genuinely different descriptions of GR, a feature which would have been missed had we applied the step-by-step Dirac procedure.\par
\indent

\section{Appendix: Commutation relations for $I_{Inst}$}
We will now infer the Poisson brackets for (\ref{CHEN88}) by inference from the corresponding canonical Ashtekar Poisson brackets
\begin{eqnarray}
\label{RELATIONS}
\{A^a_i(x),\widetilde{\sigma}^j_b(y)\}=\delta^a_b\delta^j_i\delta^{(3)}(\textbf{x},\textbf{y})
\end{eqnarray}
\noindent
along with the vanishing brackets
\begin{eqnarray}
\label{RELATIONN}
\{A^a_i(x),A^b_j(y)\}=\{\widetilde{\sigma}^i_a(x),\widetilde{\sigma}^j_b(x)\}=0.
\end{eqnarray}
\noindent
To find the analogue of (\ref{RELATIONS}) and (\ref{RELATIONN}) for (\ref{CHEN88}), we will use the tranformation equation
\begin{eqnarray}
\label{CEEDEEE}
\widetilde{\sigma}^i_a=\Psi_{ae}B^i_e,
\end{eqnarray}
\noindent
which corresponds to a noncanonical transformation.  Substitution of (\ref{CEEDEEE}) into (\ref{RELATIONS}) yields
\begin{eqnarray}
\label{COMMUTE1}
\{A^a_i(x),\Psi_{bf}(y)B^i_f(y)\}=\delta^j_i\delta^a_b\delta^{(3)}(\textbf{x},\textbf{y})\nonumber\\
\{A^a_i(x),\Psi_{bf}(y)\}B^j_f(y)+\Psi_{bf}(x)\{A^a_i(x),B^j_f(y)\}.
\end{eqnarray}
\noindent
The second term on the right hand side of (\ref{COMMUTE1}) vanishes on account of the first relations of (\ref{RELATIONN}), and upon multiplying (\ref{COMMUTE1}) by the inverse magnetic field $(B^{-1})^e_i$, assumed to be nondegenerate, we obtain
\begin{eqnarray}
\label{COMMUTE2}
\{A^a_i(x),\Psi_{bf}(y)\}=\delta^a_b(B^{-1}(y))^f_i\delta^{(3)}(\textbf{x},\textbf{y}).
\end{eqnarray}
\noindent
This gives us the Poisson brackets $\{A,A\}\sim0$ and $\{A,\Psi\}\sim{B}^{-1}$, which leaves remaining the brackets $\{\Psi,\Psi\}$.  To obtain these, we substitute (\ref{CEEDEEE}) into the second equation of (\ref{RELATIONN}), yielding 
\begin{eqnarray}
\label{COMMUTE4}
\{\widetilde{\sigma}^i_a(x),\widetilde{\sigma}^b_j(y)\}=\{\Psi_{ae}(x)B^i_e(x),\Psi_{bf}(y)B^j_f(y)\}\nonumber\\
=\Psi_{ae}(x)\{B^i_e(x),\Psi_{bf}(y)\}B^j_f(y)+\{\Psi_{ae}(x),\Psi_{bf}(y)\}B^i_e(x)B^j_f(y)\nonumber\\
+\Psi_{bf}(x)\Psi_{ae}(x)\{B^i_e(x),B^j_f(y)\}+\Psi_{bf}(y)\{\Psi_{ae}(x),B^j_f(y)\}B^i_e(x)=0.
\end{eqnarray}
\noindent
Noting that the third term vanishes on account of the first equation of (\ref{RELATIONN}), equation (\ref{COMMUTE4}) reduces to
\begin{eqnarray}
\label{COMMUTE5}
\{\Psi_{ae}(x),\Psi_{bf}(y)\}B^i_e(x)B^j_f(y)\nonumber\\
+\Psi_{ae}(x)\{B^i_e(x),\Psi_{bf}(y)\}B^j_f(y)-\Psi_{bf}(y)\{B^j_f(y),\Psi_{ae}(x)\}B^i_e(x)=0.
\end{eqnarray}
\noindent
The bottom two terms of (\ref{COMMUTE5}) can be computed using (\ref{COMMUTE2})
\begin{eqnarray}
\label{COMMUTE6}
\{B^i_e(x),\Psi_{bf}(y)\}=\epsilon^{imn}D^x_m\{A^e_n(x),\Psi_{bf}(y)\}=\epsilon^{imn}D^x_m(\delta^e_b(B^{-1}(y))^f_n\delta^{(3)}(\textbf{x},\textbf{y})).
\end{eqnarray}
\noindent
Substituting (\ref{COMMUTE6}) into (\ref{COMMUTE5}) and cancelling a pair of magnetic fields, then we have that
\begin{eqnarray}
\label{COMMUTE7}
\{\Psi_{ae}(x),\Psi_{bf}(y)\}B^i_e(x)B^j_f(y)
=\epsilon^{ijm}\bigl(\Psi_{ae}(x)D_m^x+\Psi_{ba}(y)D^y_m\bigr)\delta^{(3)}(\textbf{x},\textbf{y}).
\end{eqnarray}
\noindent
Left and right multiplying (\ref{COMMUTE7}) by the inverse of the magnetic fields, we have
\begin{eqnarray}
\label{COMMUTE8}
\{\Psi_{ae}(x),\Psi_{bf}(y)\}
=\epsilon^{ijm}\Bigl((B^{-1}(y))^f_jD^x_m\Psi_{ab}(x)(B^{-1}(x))^e_i\nonumber\\
+(B^{-1}(x))^e_iD^y_m\Psi_{ba}(y)(B^{-1}(y))^f_j\Bigr)\delta^{(3)}(\textbf{x},\textbf{y}).
\end{eqnarray}
\noindent
One sees that the internal components of $\Psi_{ae}$ have nontrivial commutation relations with themselves.\par
\indent

\end{document}